\documentclass[letter]{aa}  
\usepackage{ulem}
\usepackage{multirow}
\usepackage{graphicx}
\usepackage{txfonts}
\usepackage[colorlinks=true,linkcolor=red,anchorcolor=blue,citecolor=blue,filecolor=black,menucolor=black,runcolor=black,urlcolor=blue]{hyperref}
\usepackage{float}

\usepackage{graphicx}
\usepackage{tabularx}
\usepackage{multirow}
\usepackage{gensymb}
\usepackage{color}
\usepackage{txfonts}
\usepackage{natbib}
\usepackage{amsmath}
\usepackage{ulem}
\usepackage[flushleft]{threeparttable}
\usepackage[dvipsnames]{xcolor}
\usepackage{rotating}
\usepackage{CJKutf8}
\usepackage{makecell}

\usepackage[dvipsnames]{xcolor}

\begin{document}

   \title{Signatures of Very Massive Stars in the Epoch of Reionization}

   \author{R. Marques-Chaves\inst{\ref{inst:Geneva}}\thanks{Rui.MarquesCoelhoChaves@unige.ch}  
   \and F.~Martins\inst{\ref{inst:Montp}} 
   \and D.~Schaerer\inst{\ref{inst:Geneva},\ref{inst:CNRS}}
   \and M.~Dessauges-Zavadsky\inst{\ref{inst:Geneva}}
   \and A.~Palacios\inst{\ref{inst:Montp}}  
} 

   \institute{Geneva Observatory, Department of Astronomy, University of Geneva, Chemin Pegasi 51, CH-1290 Versoix, Switzerland \label{inst:Geneva}
   \and LUPM, Université de Montpellier, CNRS, Place Eugène Bataillon, F-34095 Montpellier, France \label{inst:Montp}
   \and CNRS, IRAP, 14 Avenue E. Belin, 31400 Toulouse, France \label{inst:CNRS}
   }

   \date{Received ; accepted}

\abstract{We present ultra-deep ($\simeq 20-30$\,hours), rest-frame UV spectroscopy with NIRSpec of two UV-bright galaxies at $z\sim 8.7$, CEERS-1019 and CEERS-1025 ($Z_{\rm neb} \simeq 0.1\,Z_{\odot}$), obtained as part of the JWST Cycle 4 SPURS large program. 
The spectra reveal exceptionally strong P-Cygni profiles in wind lines (N\,{\sc v}\,$\lambda$1240 and C\,{\sc iv}\,$\lambda$1550) and significant broad and strong He\,{\sc ii}\,$\lambda$1640 emission ($\rm EW\simeq 2-4$\,\AA). 
We compare the observations with synthetic stellar population models at $Z_{\star} \simeq 0.1\,Z_{\odot}$, both including and excluding very massive stars (VMS). Models including VMS provide a markedly improved fit to the data relative to non-VMS models ($\Delta$AIC and $\Delta$BIC $> 70$), which fail to reproduce the observed strengths of the wind features. A comparison with empirical spectra of VMS-dominated systems in the local Universe further supports this interpretation.
The best-fit VMS models imply extremely young ages of the stellar populations ($\simeq 1.5$--$2.0$\,Myr) and high ionizing photon production efficiencies ($\log \xi_{\rm ion}\,[\rm Hz\,erg^{-1}] \gtrsim 25.8$), exceeding those inferred from non-VMS models by $\sim 0.1$–$0.2$\,dex. These results provide evidence for an overabundance of VMS at high-$z$ with an IMF extending well beyond $100\,M_{\odot}$, and highlight their potential role in shaping the rest-frame UV spectra, chemical enrichment, and ionizing output of galaxies in the early Universe.
}

\titlerunning{VMS at $z\simeq 8.7$}
\maketitle

\section{Introduction}\label{Sect:intro}

Very massive stars (VMS; $M_{\star} \gtrsim 100\,M_{\odot}$; \citealt{Vink2015HiA....16...51V}) represent the extreme upper end of the stellar initial mass function (IMF). Owing to their proximity to the Eddington limit, VMS develop powerful winds, leading to high mass-loss rates ($\dot{M}$) and characteristic spectra dominated by broad emission and P-Cygni features. One of their most distinctive signatures is strong and broad He\,{\sc ii}\,$\lambda$1640 emission \citep{crowther2016, Martins2022A&A...659A.163M}. 
While such features are traditionally associated with classical Wolf-Rayet (WR) stars, VMS can produce similar spectral signatures already during core hydrogen burning (as WNh star). Their extreme luminosities and winds also make them important contributors to the ionizing photon production and chemical enrichment in young stellar populations \citep[e.g.,][]{Vink2023A&A...679L...9V, Schaerer2025A&A...693A.271S}.

VMS have been individually identified in a few young massive clusters in the Galaxy and the LMC (e.g., \citealt{Crowther2010MNRAS.408..731C}). Beyond resolved stellar populations, evidence for VMS has been reported in some local clusters and H\,{\sc ii} regions \citep[e.g.,][]{Leitherer2018ApJ...865...55L, Senchyna2021MNRAS.503.6112S, Martins2023A&A...678A.159M, Smith2023ApJ...958..194S, Wofford2023MNRAS.523.3949W}, and in galaxies at intermediate redshifts ($z\sim2-4$; \citealt{Mestric2023A&A...673A..50M, Upadhyaya2024A&A...686A.185U}), where the strength of He\,{\sc ii} emission and other features cannot be reproduced by standard synthesis models with IMFs limited to $M_{\rm up} \sim 100\,M_{\odot}$.

Despite their importance, the identification of VMS in distant galaxies remains challenging given their extremely short lifetimes ($\sim2$\,Myr). Moreover, their spectroscopic signatures can be relatively subtle in integrated spectra, often requiring high signal-to-noise (SNR) and moderate spectral resolution to disentangle them from nebular emission contributions. 
These requirements make their detection difficult at high-$z$, where deep UV spectroscopy is observationally expensive. 
Using ultra-deep rest-frame UV spectroscopy from the \textit{James Webb Space Telescope (JWST)}, we present here the first strong evidence for the presence of VMS in two UV-bright galaxies at $z\simeq 8.7$.

\section{Source description and JWST observations}\label{Sect2:data}

The two targets analyzed in this work, CEERS-1019 and CEERS-1025, are located in the EGS field and are known to reside within a large-scale overdense region and ionized bubble  at $z \simeq 8.7$ \citep{Larson2022ApJ...930..104L_enviro, Whitler2025arXiv251012019W}. 

CEERS-1019 ($\alpha$,~$\delta$ [J2000] = $215.0354^{\circ}$, $52.8907^{\circ}$; $z=8.678$) is among the UV-brightest galaxies known at $z > 8$. 
NIRSpec spectroscopy revealed intense nebular emission, including rest-UV nitrogen lines (N\,{\sc iv}] $\lambda1486$ and N\,{\sc iii}] $\lambda1750$), indicating a super-solar nitrogen abundance ($\log({\rm N/O}) \simeq -0.18$; \citealt{Marques-Chaves2024}) at relatively low metallicity ($12+\log({\rm O/H}) = 7.70$). 
A potential broad H$\beta$ component was initially interpreted as evidence for AGN activity \citep{Larson2023ApJ...953L..29L_CEERS1019}. However, deeper NIRSpec/IFS observations by \cite{Zamora2025arXiv251209022Z} revealed similar broad components in the [O\,{\sc iii}] $\lambda\lambda4960,5008$ lines, instead pointing to ionized outflows.

\begin{figure*}
  \centering
  \includegraphics[width=0.99\textwidth]{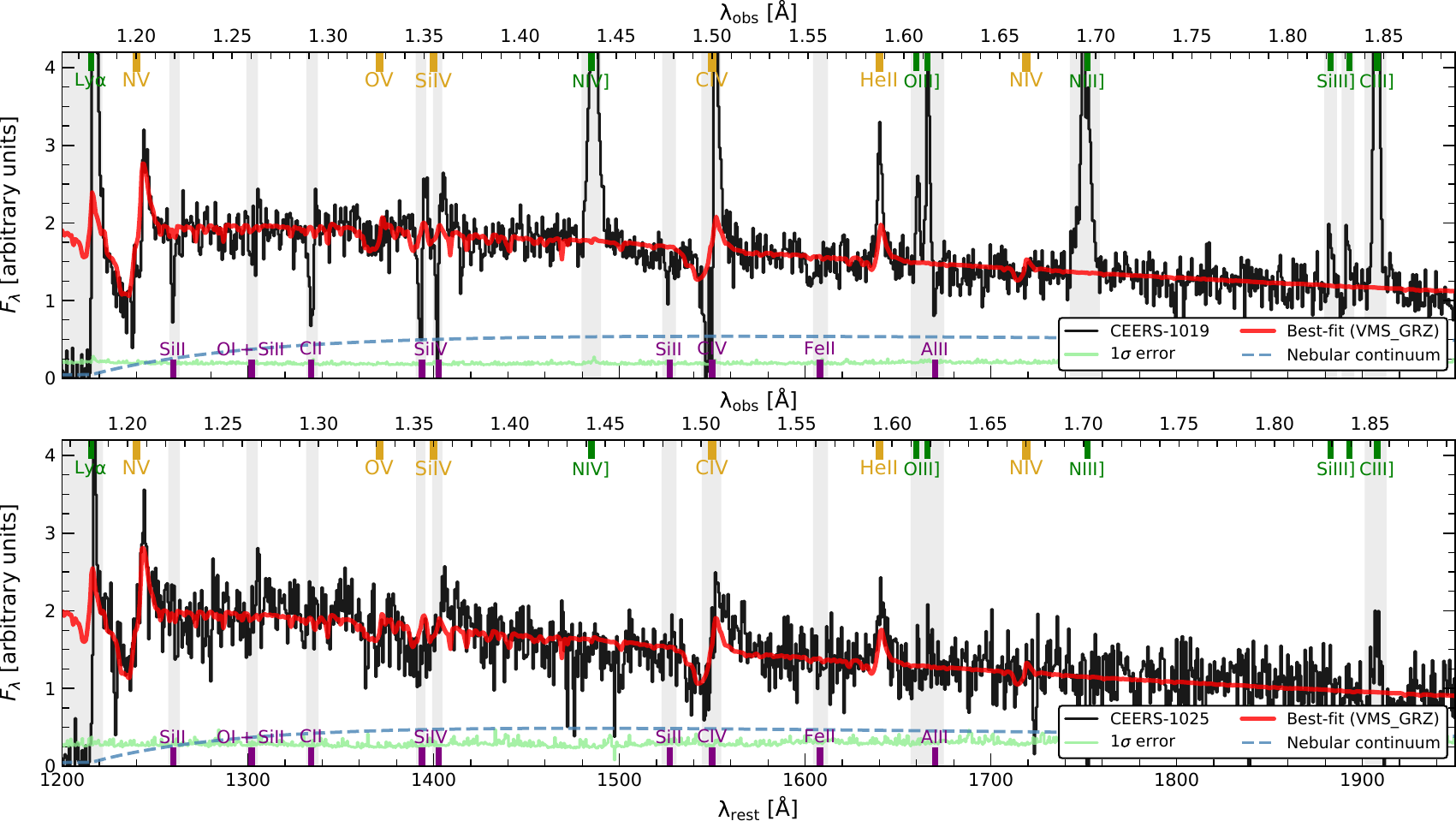}
  \caption{Rest-UV spectra of CEERS-1019 (top) and CEERS-1025 (bottom). The spectra and corresponding 1$\sigma$ uncertainties are shown in black and green, respectively. The best-fit synthetic models, which include VMS (VMS\_300\_GRZ), are overplotted in red (nebular continuum in blue). Ticks indicate the positions of selected emission and absorption features arising from nebular emission (green), stellar winds (yellow), and ISM absorption (violet). Grey regions mark spectral regions excluded from the fit due to nebular emission and ISM absorption.}
  \label{fig_spectrum}
\end{figure*}

CEERS-1025 ($\alpha$,~$\delta$ [J2000] = $214.9675^{\circ}$, $52.9330^{\circ}$; $z=8.715$) was first characterized using early CEERS observations \citep{Nakajima2023ApJS..269...33N, Tang2023MNRAS.526.1657T}. The galaxy appears young and exhibits strong rest-optical emission (\citealt{Tang2023MNRAS.526.1657T}). 
Rest-UV spectroscopy later revealed Ly$\alpha$ emission \citep{Tang2024ApJ...975..208T, Whitler2025arXiv251012019W}, along with redshifted N\,{\sc v} $\lambda1242$ emission. 
The gas-phase metallicity, $12+\log({\rm O/H}) = 7.64^{+0.18}_{-0.20}$, was derived using the $R23$ strong-line method (\citealt{Nakajima2023ApJS..269...33N}; see also: \citealt{Tang2023MNRAS.526.1657T, Tang2024ApJ...975..208T}).

Both targets were recently observed with deep rest-frame UV spectroscopy as part of the \textit{JWST} large program ``The SPectroscopic Ultra-deep Reionization-era Survey'' (SPURS; GO 9214; PIs: C. Mason \& D. Stark; \citealt{Chen2026arXiv260421516C}). Observations were obtained using the NIRSpec G140M/F100LP configuration, which provides a spectral resolution of $R \simeq 1000$ over $\lambda \simeq 0.9-1.84\,\mu$m. 
The observations employed a 3-shutter slitlet with a 3-point nodding pattern and an aperture position angle of $\mathrm{PA} \simeq 23.76^\circ$. The total exposure times were $\simeq 29.47$\,h for CEERS-1019 and $\simeq 19.64$\,h for CEERS-1025, noting that one of the slitlet shutters was closed. Level 3 data products were retrieved from MAST and reduced using the JWST Calibration Reference Data System (CRDS) version 13.0.6 with context \texttt{jwst\_1464.pmap}.

\section{Results}\label{Sect:results}

\subsection{Overview of the rest-frame UV spectra}

The high SNR spectra (Fig.~\ref{fig_spectrum}) reveal a variety of emission and absorption features arising from different components within the galaxies. Nebular emission lines are detected in both sources, including Ly$\alpha$, O\,{\sc iii}] $\lambda1666$, and C\,{\sc iii}] $\lambda1908$, as well as N\,{\sc iv}] and N\,{\sc iii}] in CEERS-1019, as discussed in previous studies \citep[e.g.,][]{Larson2023ApJ...953L..29L_CEERS1019, Marques-Chaves2024, Tang2024ApJ...975..208T, Zamora2025arXiv251209022Z}. 
The spectrum of CEERS-1019 also exhibits ISM absorption lines (violet in Fig.~\ref{fig_spectrum}), whereas these features appear weaker or absent in CEERS-1025.

More relevant for this work, remarkably strong P-Cygni features are present in both spectra, which is a key characteristic of the most massive stars and are expected in systems dominated by young stellar populations. We identify strong P-Cygni profiles in N\,{\sc v}\,$\lambda$1240 and C\,{\sc iv}\,$\lambda$1550, and slightly weaker in Si\,{\sc iv}\,$\lambda$1400. In addition, He\,{\sc ii} $\lambda1640$ emission is detected in both sources with high significance, and it appears clearly broad in CEERS-1025 and shows compelling evidence for a significant broad component in CEERS-1019 (in addition to a narrower one). In Appendix~\ref{appendix1}, we characterize the broad He\,{\sc ii} components, finding $\rm EW = 2.6\pm0.6$\,\AA\ and $\rm FWHM \simeq 1380$\,km\,s$^{-1}$ for CEERS-1019, and $\rm EW = 4.0\pm0.8$\,\AA\ and $\rm FWHM \simeq 1150$\,km\,s$^{-1}$ for CEERS-1025. Given the lack of evidence of a dominant AGN, we assume that the broad He\,{\sc ii} emission has a stellar origin.
The presence of broad He\,{\sc ii} emission, together with prominent P-Cygni profiles, indicates that the spectra are dominated by young stellar populations, potentially including VMS, which we explore next.

\subsection{Fitting the rest-UV spectra with synthetic spectral models with and without Very Massive Stars}

We use the \textsc{FiCUS} code \citep{Saldana-Lopez2023MNRAS.522.6295S} to investigate the young stellar populations of CEERS-1019 and CEERS-1025, which models observed spectra as linear combinations of simple stellar population (SSP) templates. 
We consider a suite of stellar population models, including standard \textsc{Starburst99} \citep[SB99;][]{Leitherer1999ApJS..123....3L} and \textsc{BPASS} \citep[v2.1;][]{Eldridge2017PASA...34...58E} models, with IMF upper-mass cutoffs of $120\,M_{\odot}$ and $100\,M_{\odot}$, respectively. In addition, we incorporate models from \citet{Martins2025A&A...698A.262M} that include VMS with an extended IMF up to $300\,M_{\odot}$. We explore two prescriptions for the VMS mass-loss (see: \citealt{Martins2025A&A...698A.262M}), one independent of metallicity (VMS\_GR), i.e., as those of LMC stars, and another that scales linearly with metallicity (VMS\_GRZ) relative to the LMC values. All models adopt a Salpeter-like slope at the high-mass end of the IMF. A summary of the models is provided in Table~\ref{tab:results}.

Each spectrum is fitted using SSP templates spanning ages from 0 to 40\,Myr, while the metallicity is fixed to $Z = 0.1\,Z_{\odot}$, consistent with the measured gas-phase metallicities of both sources. A nebular continuum component is included for each SSP using \textsc{PyNeb} \citep{Luridiana2015A&A...573A..42L}, assuming $n_{\rm H} = 10^{3}\,\mathrm{cm}^{-3}$ and $T_e = 1.5\times10^{4}$\,K. Dust attenuation is modeled using the \citet{Reddy2016ApJ...828..107R} law.
The best-fit solution is obtained via $\chi^2$ minimization using the \texttt{lmfit} package \citep{Newville2016ascl.soft06014N}. Uncertainties on the derived parameters are estimated through Monte Carlo simulations. 

We find that models including VMS provide significantly better fits to the spectra of both sources than standard models without VMS (BPASS or SB99), particularly in reproducing the stellar wind features (N\,{\sc v}, C\,{\sc iv}, and He\,{\sc ii}). Fig.~\ref{fig_spectrum} shows the best-fits for CEERS-1019 and CEERS-1025, both corresponding to VMS\_GRZ. The full set of best-fit models is shown in Appendix~\ref{appendix3} (Figs.~\ref{fig_appendix3} and \ref{fig_appendix4}).
Quantitatively, the Akaike (AIC) and Bayesian (BIC) information criteria strongly favor VMS models, with $\Delta \rm AIC$ and $\Delta \rm BIC$ of at least $\gtrsim 70-280$ relative to non-VMS models (Table~\ref{tab:results}). These differences indicate overwhelming statistical support for the inclusion of VMS in the modelling. A more detailed assessment, comparing the observed strengths of the He\,{\sc ii} emission and N\,{\sc v} P-Cygni profiles with model predictions, is provided in Appendix~\ref{appendix2} and Fig.~\ref{fig_appendix2}.

In particular, we find that standard models without VMS fail to reproduce several key spectral features (Figs.~\ref{fig_appendix3} and \ref{fig_appendix4}). For N\,{\sc v}, non-VMS models significantly underestimate the strength of the P-Cygni profile, especially the redshifted emission component, whereas VMS models provide an excellent match to both the absorption and emission features.
A similar trend is observed for the broad He\,{\sc ii} emission. 
At these metallicities (0.1\,$Z_{\odot}$), the contribution of classical WR stars is expected to be limited, as their $\dot{M}$ and wind strengths generally decrease with $Z$ \citep{Crowther2023MNRAS.521..585C}. Empirically, WR-dominated systems at LMC-like metallicity typically exhibit $\rm EW \sim 1$–2\,\AA\ (\citealt{Martins2023A&A...678A.159M}), suggesting that at even lower metallicities, the WR contribution could be even weaker, as predicted by the non-VMS models ($\rm EW \lesssim 1$\,\AA, Fig.~\ref{fig_appendix2}).
However, recent theoretical works suggest a more nuanced dependence of $\dot{M}$ on both metallicity and Eddington factor ($\Gamma$) in WR, with relatively high $\dot{M}$ still possible at very high $\Gamma$ \citep[e.g.][]{Sander2020MNRAS.491.4406S}. 
While VMS models substantially improve the agreement with the measured He\,{\sc ii} line strengths (Fig.~\ref{fig_appendix2}), we cannot exclude some contribution from classical WR stars, which could be tested with deep rest-optical spectroscopy (WR “bumps”; \citealt{Rivera-Thorsen2024A&A...690A.269R}). 
The C\,{\sc iv} P-Cygni profile in CEERS-1025 is also better reproduced by models including VMS (particularly the VMS\_GRZ model). 
In CEERS-1019, the comparison is less straightforward because the C\,{\sc iv} profile is strongly affected by nebular emission and ISM absorption, which complicates a detailed decomposition of the stellar wind component. Finally, CEERS-1025 exhibits a prominent Si\,{\sc iv}\,$\lambda$1400 P-Cygni profile that is not well reproduced by any of the models, although the VMS\_GRZ one appears to partially account for the observed feature. Strong Si\,{\sc iv} P-Cygni profiles are typically associated with OB supergiants \citep{Walborn1987PASP...99...40W}, suggesting somewhat cooler stellar populations. However, similar Si\,{\sc iv} features have also been reported in UV-bright VMS-dominated candidates at $z\sim 2-4$ \citep{Upadhyaya2024A&A...686A.185U}.

Based on the best-fit VMS\_GRZ models, we derive light-weighted stellar ages of $1.5\pm0.4$\,Myr and $1.9\pm0.3$\,Myr for CEERS-1019 and CEERS-1025, respectively (Table~\ref{tab:results}). 
Both sources exhibit elevated ionizing photon production efficiencies, $\xi_{\rm ion} = Q_{\rm H}/L_{\rm UV} \gtrsim 10^{25.8}$\,Hz\,erg$^{-1}$, exceeding the values inferred from models without VMS by $\gtrsim 0.1-0.2$\,dex.

\subsection{Comparison with other VMS-dominated systems}\label{comparison}

\begin{figure*}
  \centering
  \includegraphics[width=0.95\textwidth]{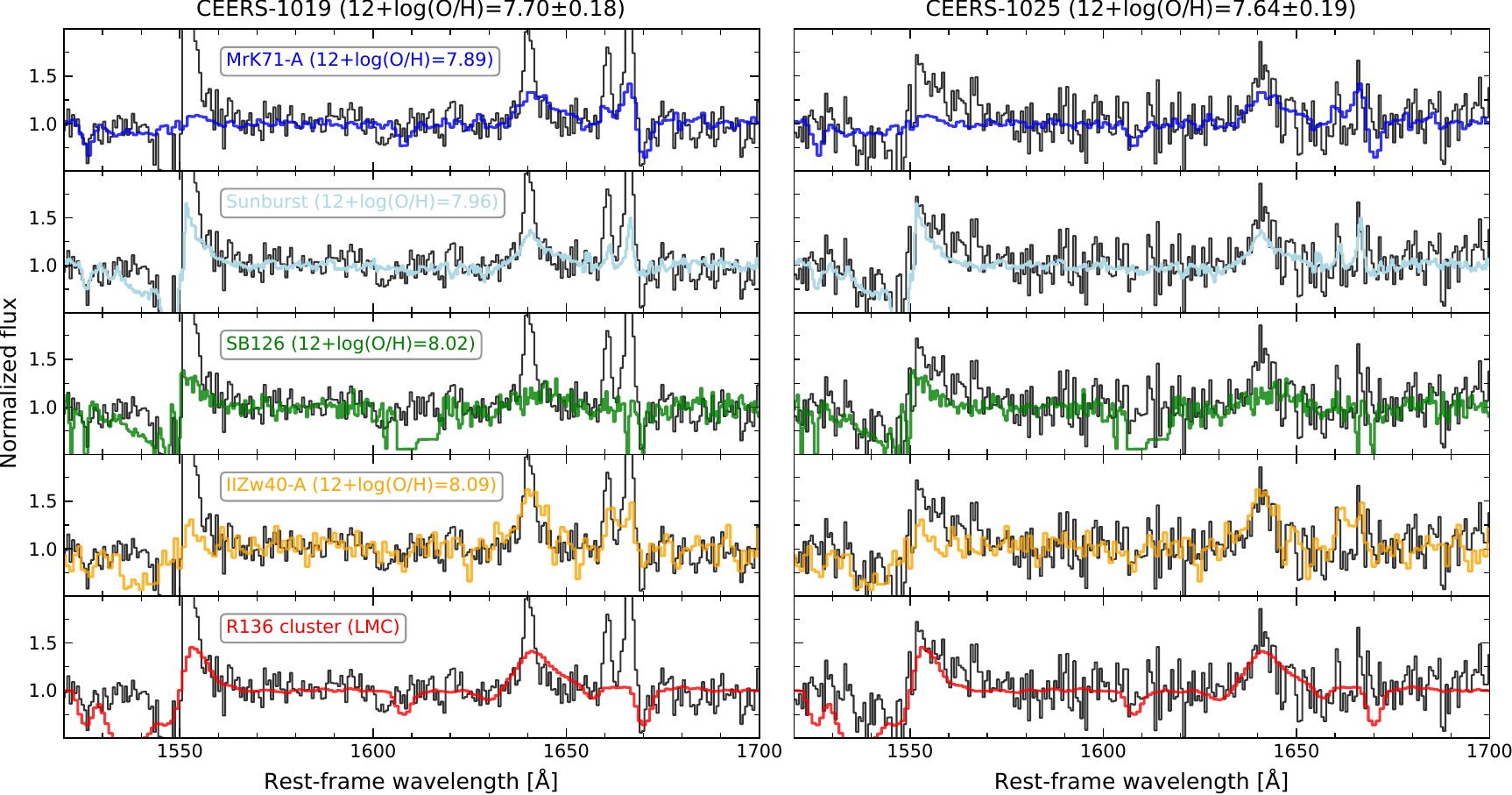}
  \caption{Comparison of the G140M spectra (black) of CEERS-1019 (left) and CEERS-1025 (right) around C\,{\sc iv} and He\,{\sc ii} with other VMS-dominated system candidates (sorted by metallicity): MrK71-A (blue), Sunburst cluster (light-blue), SB126 (green), IIZw40-A (orange), and R136 (red).}
  \label{fig_comp}
\end{figure*}

We now compare the G140M spectra of CEERS-1019 and CEERS-1025 with those of known and candidate VMS-dominated systems spanning similar or moderately higher metallicities. Fig.~\ref{fig_comp} presents the normalized spectra of both sources around C\,{\sc iv} and He\,{\sc ii}, alongside local star clusters MrK71-A ($Z/Z_{\odot}\simeq16\%$; \citealt{Smith2023ApJ...958..194S}), SB126 ($Z/Z_{\odot}\simeq21\%$; \citealt{Senchyna2021MNRAS.503.6112S}), II\,Zw\,40-A ($Z/Z_{\odot}\simeq25\%$; \citealt{Leitherer2018ApJ...865...55L}), and R136 in the LMC (\citealt{crowther2016}), as well as the highly magnified Sunburst cluster at $z=2.37$ ($Z/Z_{\odot}\simeq19\%$; \citealt{Chisholm2019ApJ...882..182C, Mestric2023A&A...673A..50M, Welch2025ApJ...980...33W}).

Despite differences in data quality and properties among the comparison samples (e.g., nebular contribution, metallicity, and age), and between the two sources themselves (with CEERS-1019 showing stronger nebular emission), the overall spectral appearance of CEERS-1019 and CEERS-1025 resembles that of some VMS-dominated systems.
In CEERS-1019, the C\,{\sc iv} P-Cygni emission is broadly consistent with that observed in the Sunburst cluster, SB126, and R136, although the profile near systemic velocity is partially contaminated by nebular emission. In contrast, its relatively weak blueshifted absorption is more similar to that seen in MrK71-A. Excluding its central (nebular) component, the broad wings of He\,{\sc ii} are also consistent with those of the Sunburst cluster. For CEERS-1025, the C\,{\sc iv} profile shows similarities to those of the Sunburst and SB126 clusters. However, its He\,{\sc ii} emission more closely resembles that of the most metal-rich sources, II\,Zw\,40-A and R136, which exhibit stronger He\,{\sc ii} (EW $\gtrsim 4$\,\AA).

\section{Discussion and Conclusions}\label{Sect4:disc}

Based on spectral modeling and comparisons with local VMS-dominated systems, our results provide strong evidence that CEERS-1019 and CEERS-1025 ($z \simeq 8.7$; $Z/Z_{\odot}\simeq0.1$) host a significant population of VMS. We explore two assumptions for the VMS $\dot{M}$–$Z$ dependence. In one case (VMS\_GR), mass-loss rates are the same as those of LMC stars; in the other (VMS\_GRZ), we adopt a linear scaling with $Z$ relative to LMC values (see: \citealt{Martins2025A&A...698A.262M}). Both VMS models provide significantly improved fits compared to models without VMS ($\Delta$AIC and $\Delta$BIC $> 70$). However, we find no significant differences between the two prescriptions, and thus cannot constrain how VMS mass-loss rates depend on metallicity. Furthermore, while the VMS contribution appears robust, the IMF $M_{\rm up} = 300\,M_{\odot}$ assumed here should be regarded as indicative, given degeneracies between the IMF slope and the $\dot{M}$–$Z$ dependence.

An additional source of uncertainty is the stellar metallicity adopted in the models. While we fixed the stellar metallicity to $Z_{\star}/Z_{\odot}=0.1$ based on the nebular one ($12+\log({\rm O/H}) \simeq 7.70$), the exact stellar metallicities of these systems remain uncertain.  Nevertheless, the presence of VMS in these sources appear robust against plausible metallicity variations. The observed He\,{\sc ii} and N\,{\sc v} P-Cygni strengths are reproduced by VMS models spanning $Z_{\star}/Z_{\odot}=0.1$--$0.2$ (Fig.~\ref{fig_appendix2}), while current non-VMS population synthesis models fail to reproduce the observed He\,{\sc ii} strengths even at LMC metallicity \citep{Martins2023A&A...678A.159M}. On the other hand, if the stellar metallicity is lower than the nebular one, as expected in $\alpha$-enhanced systems enriched predominantly by CCSNe, even more extreme stellar populations would likely be required (e.g. top-heavy IMFs and/or larger IMF $M_{\rm up}$).

The presence of VMS in sources at the epoch of reionization is somewhat expected, as they are observed in the Milky Way and LMC and are likely present in other nearby systems, while their formation is predicted to be favored in the dense, low-metallicity environments typical of high-$z$ galaxies. However, their identification remains challenging due to their short lifetimes ($\sim$2\,Myr). Whether VMS are ubiquitous in the early Universe or instead trace only a subset of extreme systems thus remains an open question. If common, as in the $z\,\sim\,2$--$4$ UV-bright galaxies studied by \cite{Upadhyaya2024A&A...686A.185U}, they could significantly enhance both the UV luminosities and the ionizing photon production efficiencies, as inferred here ($\xi_{\rm ion} \gtrsim 10^{25.8}\,\mathrm{Hz\,erg}^{-1}$).

Furthermore, CEERS-1019 belongs to the class of nitrogen-enhanced galaxies recently identified at high-$z$. Several scenarios have been proposed to explain their extreme N/O ratios, including enrichment from WRs/massive stars, VMS, or supermassive stars (SMS), among others. While our results could support a VMS contribution \citep{Vink2023A&A...679L...9V}, no firm conclusion can be drawn from a single object. Moreover, these scenarios are not mutually exclusive since a young VMS-dominated population could outshine previous star-formation episodes hosting WR, while the presence of VMS would also be naturally expected if the IMF upper-mass limit extends into the SMS regime ($>10^{3}\,M_{\odot}$).

More broadly, these results highlight the importance of deep rest-frame UV spectroscopy and the role of VMS in shaping the spectra of young galaxies during the epoch of reionization. Upcoming deeper rest-UV spectroscopy with \textit{JWST} and soon with the ELT, will be essential to establish statistically whether VMS and different IMFs are common in typical high-redshift star-forming galaxies, and to better understand their impact on stellar evolution, chemical enrichment, and, more generally, their role in early galaxy evolution and cosmic reionization.

\bibliographystyle{aa} 
\bibliography{bibliography.bib}

\onecolumn

\begin{appendix}

\section{Acknowledgements}

\begin{acknowledgements}

The authors thank the referee for useful comments. This work is based on observations made with the NASA/ESA/CSA James Webb Space Telescope. The data were obtained from the Mikulski Archive for Space Telescopes at the Space Telescope Science Institute, which is operated by the Association of Universities for Research in Astronomy, Inc., under NASA contract NAS 5-03127 for JWST. These observations are associated with program \#9214.
The authors acknowledge the SPURS team (led by coPIs C. Masond and D. Stark) for developing their observing program with a zero-exclusive-access period.
The data described here may be obtained from \url{https://dx.doi.org/10.17909/tb65-jv89}.

\end{acknowledgements}

\section{Decomposition of the He\,{\sc ii} $\lambda$1640 emission}\label{appendix1}

\begin{figure*}[h!]
\centering
  \includegraphics[width=0.95\textwidth]{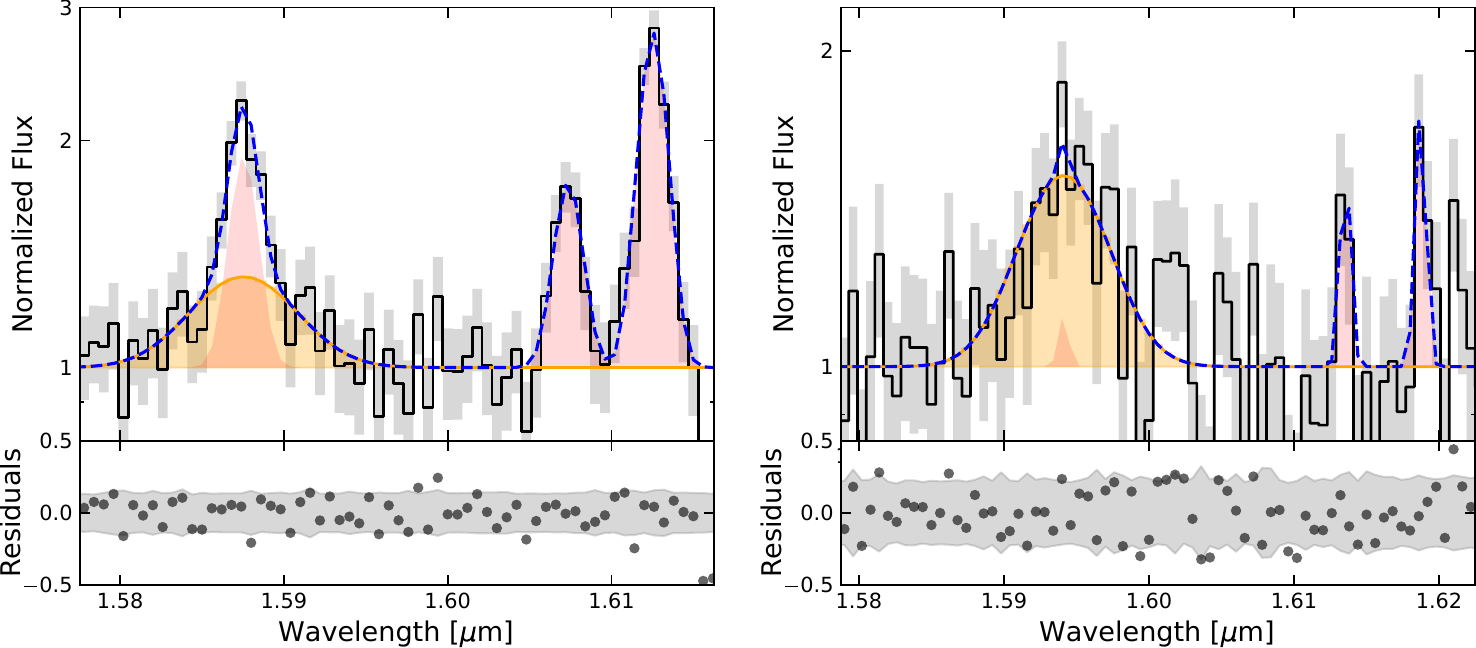}
    \caption{Normalized spectra (black) of CEERS-1019 (left) and CEERS-1025 (right) in the spectral region around He\,{\sc ii} and O\,{\sc iii}] $\lambda\lambda1660,1666$. The top panels show the best-fit models (dashed blue), where the He\,{\sc ii} profile is decomposed into two Gaussian components: a narrow component with its width fixed to that of the O\,{\sc iii}] lines (shaded red), and a broader component (orange). The bottom panels show the residuals of the fits.}
  \label{fig_appendix1}
\end{figure*}

We investigate the He\,{\sc ii} emission-line profiles in CEERS-1019 and CEERS-1025. 
As a first step, we fit the He\,{\sc ii} and O\,{\sc iii}] $\lambda\lambda1660,1666$ emission lines using a single Gaussian component for each line. We find that the O\,{\sc iii}] lines are well described by narrow Gaussians, with $\rm FWHM\,(OIII]) = 380\pm140$\,km\,s$^{-1}$ and $148\pm72$\,km\,s$^{-1}$ for CEERS-1019 and CEERS-1025, respectively. 
In contrast, the He\,{\sc ii} emission is significantly broader in both sources, with $\rm FWHM = 607\pm90$\,km\,s$^{-1}$ and $1154\pm301$\,km\,s$^{-1}$ for CEERS-1019 and CEERS-1025, respectively. Moreover, in CEERS-1019 the single-Gaussian model fails to reproduce the extended wings of the He\,{\sc ii} profile (Fig.~\ref{fig_appendix1}), further indicating the presence of an additional broad component.

We therefore model the He\,{\sc ii} profile in CEERS-1019 using two Gaussian components: a narrow component, with its width fixed to that of the nebular O\,{\sc iii}] emission ($\rm FWHM\,(OIII]) = 380\pm140$\,km\,s$^{-1}$), and a broader component with free width. For the latter, we measure $\rm FWHM\,(HeII) = 1373\pm392$\,km\,s$^{-1}$ and $\rm EW\,(HeII) = 2.58\pm0.54$\,\AA, contributing $\simeq 57\%$ of the total He\,{\sc ii} flux.
In contrast, the He\,{\sc ii} profile in CEERS-1025 is well described by a single broad Gaussian component, with no significant evidence for a narrow (nebular) contribution. In this case, we derive $\rm FWHM\,(HeII) = 1156\pm450$\,km\,s$^{-1}$ and $\rm EW\,(HeII) = 3.99\pm0.64$\,\AA.

Fig.~\ref{fig_appendix1} presents the best-fit Gaussian models for the He\,{\sc ii} and O\,{\sc iii}] emission lines in both sources.

\section{Best-fit synthetic models with and without Very Massive Stars}\label{appendix3}

\begin{figure*}[h!]
\centering
  \includegraphics[width=0.98\textwidth]{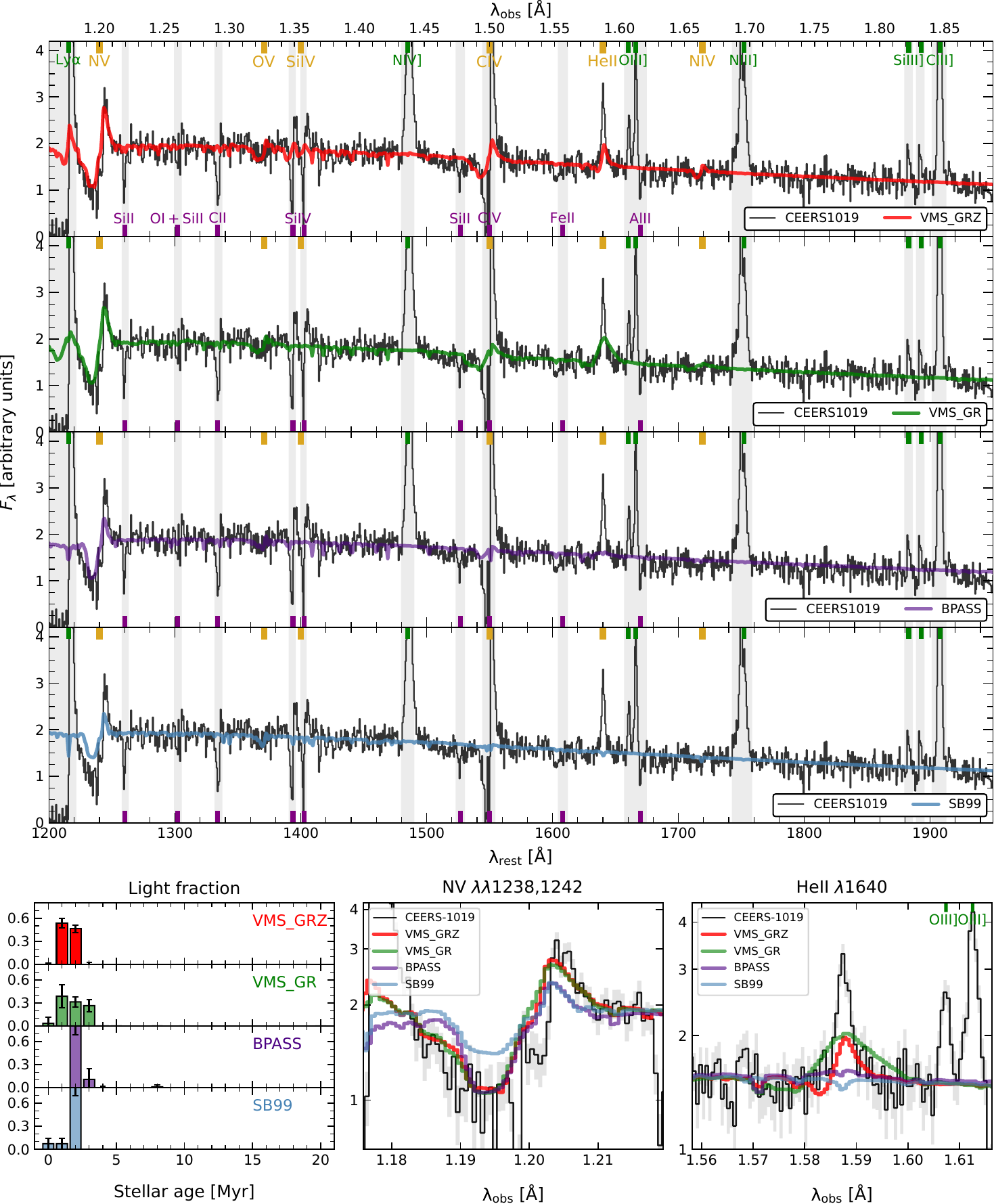}
  \caption{Rest-frame UV spectrum of CEERS-1019 (black) and best-fit synthetic models derived with \textsc{FiCUS}: VMS\_300\_GRZ (red), VMS\_300\_GR (green), BPASS (violet), and SB99 (blue). Grey regions indicate wavelength intervals excluded from the fit. The lower panels show the light-weighted contributions of the SSP components (left), and zoom-ins on the N\,{\sc v}\,$\lambda1240$ (middle) and He\,{\sc ii}\,$\lambda1640$ (right) spectral regions.}
  \label{fig_appendix3}
\end{figure*}

\begin{figure*}[h!]
\centering
  \includegraphics[width=0.98\textwidth]{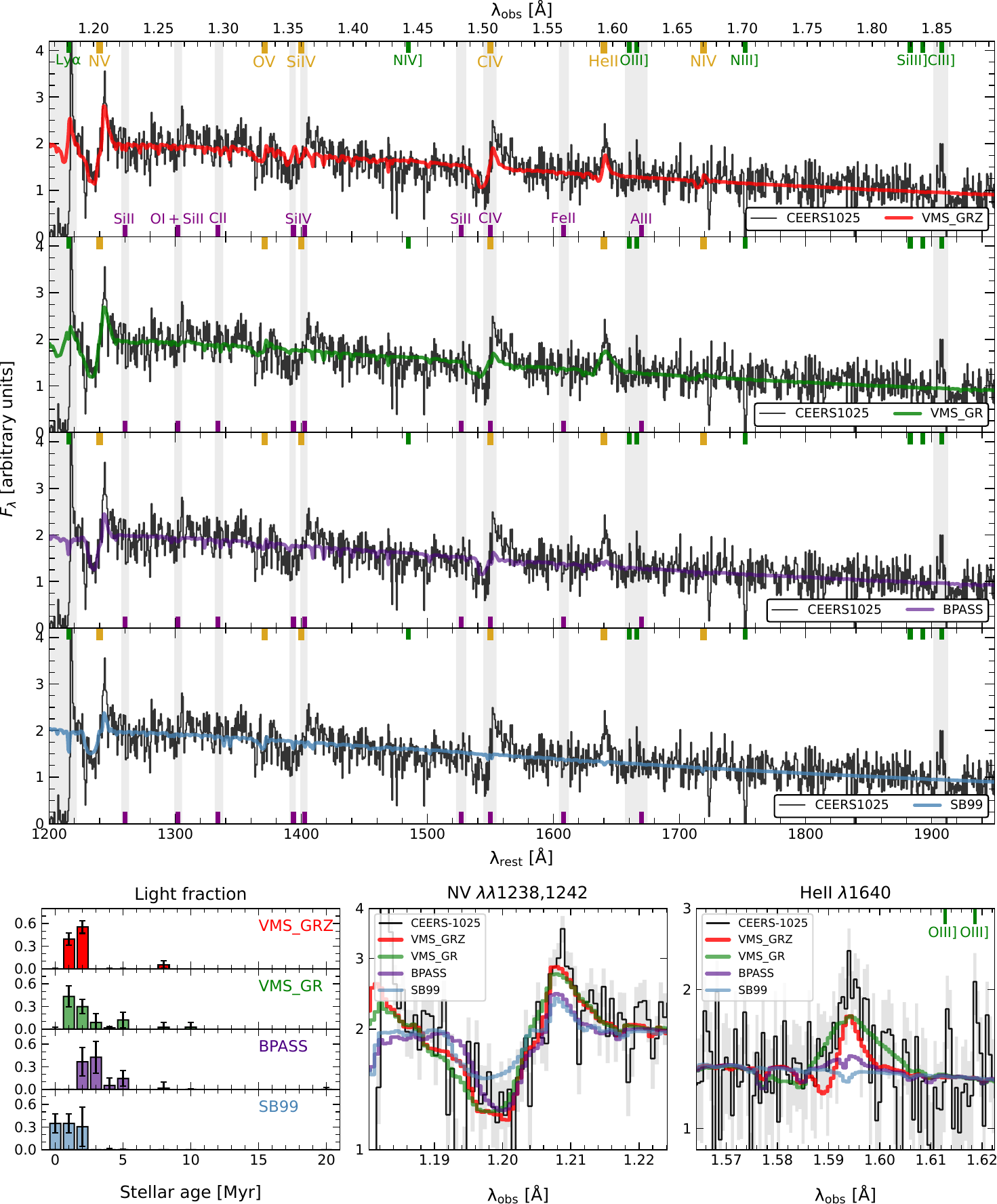}
  \caption{Same as Fig.~\ref{fig_appendix3}, but for CEERS-1025.}
  \label{fig_appendix4}
\end{figure*}

Figs.~\ref{fig_appendix3} and \ref{fig_appendix4} present the best-fit synthetic spectra for CEERS-1019 and CEERS-1025, respectively, obtained using \textsc{FiCUS} \citep{Saldana-Lopez2023MNRAS.522.6295S} for models at $Z=0.1\,Z_{\odot}$ including VMS (VMS\_GRZ and VMS\_GR; shown in red and green) and without VMS (BPASS and SB99; shown in violet and blue). Table~\ref{tab:results} summarizes the statistical assessment and best-fit parameters for each model.

\begin{table*}
\begin{center}
\caption{Summary of models, statistical assessment, and best-fit parameters obtained for CEERS-1019 and CEERS-1025.}
\label{tab:results}
\begin{tabular}{llcccccccc}
\hline \hline
 &  &  &  & \multicolumn{3}{c}{Statistical Assessment} & \multicolumn{3}{c}{Best-fit Parameters} \\
Source & Model ID & IMF $M_{\rm up}$ & $\dot{M}$--$Z$ scale & $\chi^2_{\rm red}$ & AIC & BIC & Age & $E(B-V)$ & log($\xi_{\rm ion})$  \\
  &  [$Z/Z_{\odot}=10\%$] &  [$M_{\odot}$]  &   &   &   &   & [Myr]  & [mag.]  & $\rm [Hz\, erg^{-1}]$  \\  
\hline 

\multirow{4}{*}{\makecell[l]{CEERS-1019 \\ {\small $z=8.678$} \\ {\small $M_{\rm UV} = -22.4$} }}
& VMS\_GRZ & 300 & yes & 1.25 & 1296 & 1247 & $1.46 \pm 0.35$ & $0.18 \pm 0.03$ & $25.84 \pm 0.02$ \\
 & VMS\_GR  & 300 & no & 1.26 & 1308 & 1259 & $1.81 \pm 0.17$ & $0.19\pm0.02$ & $25.81\pm0.02$ \\
 & BPASS    & 100  & --- & 1.47 & 1512 & 1463 & $2.17 \pm 0.18$ & $0.25\pm0.02$ & $25.71\pm0.02$ \\
 & SB99     & 120  & --- & 1.54 & 1583 & 1534 & $1.78\pm0.22$ & $0.23 \pm 0.01$ & $25.69 \pm 0.02$ \\

\hline 

\multirow{4}{*}{\makecell[l]{CEERS-1025 \\ {\small $z=8.715$} \\ {\small $M_{\rm UV} = -21.2$} }}
 & VMS\_GRZ & 300 & yes & 1.25 & 1377 & 1337 & $1.94\pm0.31$ & $0.08\pm0.02$ & $25.81\pm0.02$ \\
 & VMS\_GR  & 300 & no & 1.26 & 1385 & 1340 & $2.39\pm0.41$ & $0.10 \pm 0.02$ & $25.79\pm0.03$ \\
 & BPASS    & 100  & --- & 1.33 & 1464 & 1414 & $3.04\pm0.54$ & $0.12\pm0.02$ & $25.63\pm0.04$ \\
 & SB99     & 120  & --- & 1.35 & 1472 & 1422 & $0.97 \pm 0.37$ & $0.13\pm0.02$ & $25.70 \pm 0.02$ \\

\hline \hline
\end{tabular}
\end{center}
\end{table*}

\section{Strength of the He\,{\sc ii} and N\,{\sc v} P-Cygni profiles and comparison with synthetic stellar models}\label{appendix2}

We measure the rest-frame equivalent widths of the He\,{\sc ii} emission and the N\,{\sc v} P-Cygni profile for CEERS-1019 and CEERS-1025, as well as for the suite of models considered in this work (see also \citealt{Upadhyaya2024A&A...686A.185U}). 
To this end, we normalize the spectra by fitting a linear continuum using line-free regions around He\,{\sc ii} ($\lambda_{\rm rest} = 1610-1620$\,\AA\ and $1651-1658$\,\AA) and N\,{\sc v} ($\lambda_{\rm rest} = 1265-1295$\,\AA), ensuring that strong spectral features do not bias the continuum determination. The He\,{\sc ii} EW is measured over the interval $\lambda_{\rm rest}=1633-1647$\,\AA, finding $\rm EW\,(HeII) = 4.53\pm0.36$\,\AA\ and $4.09\pm0.71$\,\AA\ for CEERS-1019 and CEERS-1025, respectively. For the N\,{\sc v} P-Cygni profile, we measure the absorption and emission components separately using the windows $\lambda_{\rm rest}=1225-1239$\,\AA\ and $1240-1253$\,\AA, respectively.

\begin{figure*}[h!]
\centering
  \includegraphics[width=0.80\textwidth]{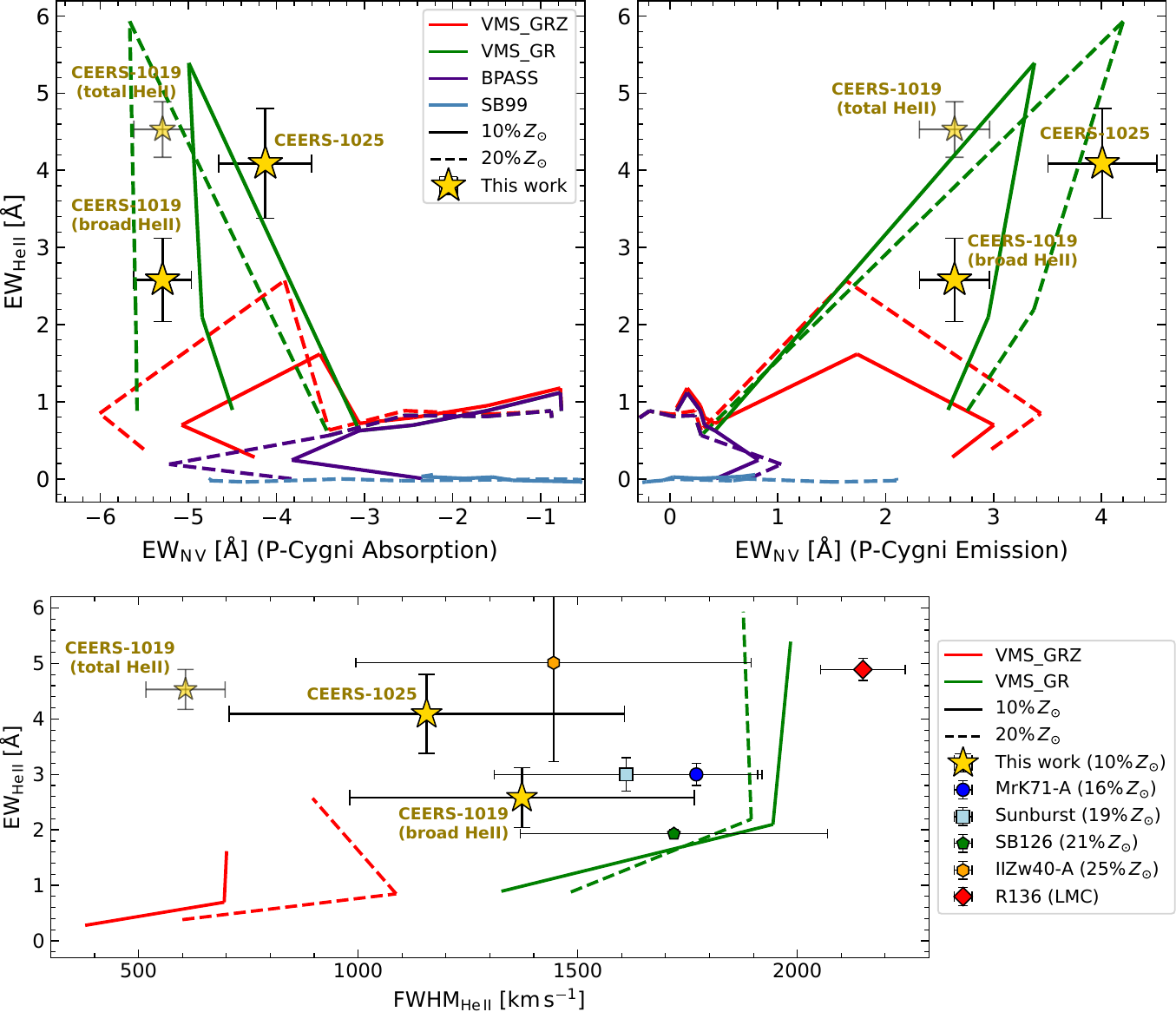}
  \caption{Relationship between the strength of the He\,{\sc ii} emission and the N\,{\sc v} P-Cygni absorption (top left) and emission (top right), as predicted by different models. Models including VMS are shown in red and green, while those without VMS are shown in violet and blue. Solid and dashed lines correspond to metallicities of $Z/Z_{\odot}=10\%$ and $20\%$, respectively. Measurements for CEERS-1019 and CEERS-1025 are indicated by stars, except for the broad He\,{\sc ii} component in CEERS-1019, which is derived from Gaussian decomposition. The bottom panel shows the strength of the He\,{\sc ii} emission as a function of its line width (FWHM), for VMS models (lines) and for  CEERS-1019 and CEERS-1025 (stars) and other VMS-dominated systems and candidates: MrK71-A (blue), Sunburst cluster (light-blue), SB126 (green), IIZw40-A (orange), and R136 (red).}
  \label{fig_appendix2}
\end{figure*}

Top panels of Fig.~\ref{fig_appendix2} shows the relationship between the EWs of He\,{\sc ii} and the N\,{\sc v} absorption (left) and emission (right) components. The measurements for CEERS-1019 and CEERS-1025 (stars) are compared with predictions from SSP models spanning a range of ages and metallicities ($Z/Z_{\odot}=10\%$ and $20\%$, shown as solid and dashed lines, respectively). 
We find that only models including VMS (red and green) can simultaneously reproduce the observed strengths of both the N\,{\sc v} P-Cygni emission and the He\,{\sc ii} line. In contrast, models without VMS fail to match the data. BPASS models, which include binary evolution, can reach at most $\rm EW(HeII) \simeq 1$\,\AA, driven by the contribution of Wolf-Rayet (WR) stars, but only at ages of $\gtrsim 5-8$\,Myr, when the N\,{\sc v} feature has already significantly weakened. SB99 models, on the other hand, predict negligible He\,{\sc ii} emission.

The bottom panel of Fig.~\ref{fig_appendix2} shows the relationship between the EWs of He\,{\sc ii} emission line and its line width (FWHM) for the VMS models (lines) and CEERS-1019 and CEERS-1025 (stars). We also show the measurements of other VMS-dominated systems and candidates discussed in Sect.~\ref{comparison}: MrK71-A (blue), Sunburst cluster (light-blue), SB126 (green), IIZw40-A (orange), and R136 (red).

\end{appendix}

\end{document}